\documentclass[aps,prl,superscriptaddress,altaffilletter,floatfix,preprintnumbers,amsmath,amssymb,amsbsy,nofootinbib,nobibnotes,twocolumn,showpacs]{revtex4-1}

\usepackage{graphicx,color,units,bm,subfigure}

\newcommand{\svnid}[1]{ } 

\newcommand{\macro}[1]{\textcolor{black}{#1}} 

\newcommand\CBCEVENTSIGMA{\macro{\ensuremath{5.1}}}

\newcommand{\ninetyrange}[3]{\ensuremath{{#1}^{+{#2}}_{-{#3}}}}

\newcommand{\pergpcyr}{\ensuremath{\mathrm{Gpc}^{-3}\,\mathrm{yr}^{-1}}}

\newcommand{\alphabetrateone}{\macro{\ensuremath{\ninetyrange{16}{38}{13} \, \pergpcyr}}}

\newcommand{\DISTANCECOMPACT}{\macro{\ensuremath{410_{-180}^{+160}}}} 
\newcommand{\MONESCOMPACT}{\macro{\ensuremath{36_{-4}^{+5}}}} 
\newcommand{\MTWOSCOMPACT}{\macro{\ensuremath{29_{-4}^{+4}}}} 
\newcommand{\REDSHIFTCOMPACT}{\macro{\ensuremath{0.09_{-0.04}^{+0.03}}}} 
\newcommand{{\NCYCLES}}{{\macro{{\ensuremath{{10}}}}}} 

\newcommand{\alphabetrateoneSIMPLE}{\ensuremath{16}}

\newcommand{\MCSOne}{\ensuremath{28}} 
\newcommand{\spinOne}{\ensuremath{-0.06}} 
\newcommand{\OmegaRange}{\ensuremath{1.1_{-0.9}^{+2.7} \times 10^{-9} }}

\newcommand{\Fiducial}{\texttt{Fiducial}}
\newcommand{\AltSFR}{\texttt{AltSFR}}
\newcommand{\ConstRate}{\texttt{ConstRate}}
\newcommand{\LongDelay}{\texttt{LongDelay}}
\newcommand{\LowMetallicity}{\texttt{LowMetallicity}}
\newcommand{\FlatDelay}{\texttt{FlatDelay}}
\newcommand{\TwoEvents}{\texttt{LowMass}}

\begin{document}

\title{GW150914: Implications for the stochastic gravitational-wave background from binary black holes}
\date{\today}
\author{%
B.~P.~Abbott,$^{1}$  
R.~Abbott,$^{1}$  
T.~D.~Abbott,$^{2}$  
M.~R.~Abernathy,$^{1}$  
F.~Acernese,$^{3,4}$
K.~Ackley,$^{5}$  
C.~Adams,$^{6}$  
T.~Adams,$^{7}$
P.~Addesso,$^{3}$  
R.~X.~Adhikari,$^{1}$
% ...
\\
\medskip
(LIGO Scientific Collaboration and Virgo Collaboration) 
\\
\medskip
{{}$^{\dag}$Deceased, May 2015. {}$^{\ddag}$Deceased, March 2015. }%
}\noaffiliation
\affiliation {LIGO, California Institute of Technology, Pasadena, CA 91125, USA }
\affiliation {Louisiana State University, Baton Rouge, LA 70803, USA }
\affiliation {Universit\`a di Salerno, Fisciano, I-84084 Salerno, Italy }
\affiliation {INFN, Sezione di Napoli, Complesso Universitario di Monte S.Angelo, I-80126 Napoli, Italy }
\affiliation {University of Florida, Gainesville, FL 32611, USA }
\affiliation {LIGO Livingston Observatory, Livingston, LA 70754, USA }
\affiliation {Laboratoire d'Annecy-le-Vieux de Physique des Particules (LAPP), Universit\'e Savoie Mont Blanc, CNRS/IN2P3, F-74941 Annecy-le-Vieux, France }
\affiliation {Albert-Einstein-Institut, Max-Planck-Institut f\"ur Gravi\-ta\-tions\-physik, D-30167 Hannover, Germany }
\affiliation {Nikhef, Science Park, 1098 XG Amsterdam, The Netherlands }
\affiliation {LIGO, Massachusetts Institute of Technology, Cambridge, MA 02139, USA }
\affiliation {Instituto Nacional de Pesquisas Espaciais, 12227-010 S\~{a}o Jos\'{e} dos Campos, SP, Brazil }
\affiliation {INFN, Gran Sasso Science Institute, I-67100 L'Aquila, Italy }
\affiliation {INFN, Sezione di Roma Tor Vergata, I-00133 Roma, Italy }
\affiliation {Inter-University Centre for Astronomy and Astrophysics, Pune 411007, India }
\affiliation {International Centre for Theoretical Sciences, Tata Institute of Fundamental Research, Bangalore 560012, India }
\affiliation {University of Wisconsin-Milwaukee, Milwaukee, WI 53201, USA }
\affiliation {Leibniz Universit\"at Hannover, D-30167 Hannover, Germany }
\affiliation {Universit\`a di Pisa, I-56127 Pisa, Italy }
\affiliation {INFN, Sezione di Pisa, I-56127 Pisa, Italy }
\affiliation {Australian National University, Canberra, Australian Capital Territory 0200, Australia }
\affiliation {The University of Mississippi, University, MS 38677, USA }
\affiliation {California State University Fullerton, Fullerton, CA 92831, USA }
\affiliation {LAL, Univ. Paris-Sud, CNRS/IN2P3, Universit\'e Paris-Saclay, Orsay, France }
\affiliation {Chennai Mathematical Institute, Chennai, India }
\affiliation {Universit\`a di Roma Tor Vergata, I-00133 Roma, Italy }
\affiliation {University of Southampton, Southampton SO17 1BJ, United Kingdom }
\affiliation {Universit\"at Hamburg, D-22761 Hamburg, Germany }
\affiliation {INFN, Sezione di Roma, I-00185 Roma, Italy }
\affiliation {Albert-Einstein-Institut, Max-Planck-Institut f\"ur Gravitations\-physik, D-14476 Potsdam-Golm, Germany }
\affiliation {APC, AstroParticule et Cosmologie, Universit\'e Paris Diderot, CNRS/IN2P3, CEA/Irfu, Observatoire de Paris, Sorbonne Paris Cit\'e, F-75205 Paris Cedex 13, France }
\affiliation {Montana State University, Bozeman, MT 59717, USA }
\affiliation {Universit\`a di Perugia, I-06123 Perugia, Italy }
\affiliation {INFN, Sezione di Perugia, I-06123 Perugia, Italy }
\affiliation {European Gravitational Observatory (EGO), I-56021 Cascina, Pisa, Italy }
\affiliation {Syracuse University, Syracuse, NY 13244, USA }
\affiliation {SUPA, University of Glasgow, Glasgow G12 8QQ, United Kingdom }
\affiliation {LIGO Hanford Observatory, Richland, WA 99352, USA }
\affiliation {Wigner RCP, RMKI, H-1121 Budapest, Konkoly Thege Mikl\'os \'ut 29-33, Hungary }
\affiliation {Columbia University, New York, NY 10027, USA }
\affiliation {Stanford University, Stanford, CA 94305, USA }
\affiliation {Universit\`a di Padova, Dipartimento di Fisica e Astronomia, I-35131 Padova, Italy }
\affiliation {INFN, Sezione di Padova, I-35131 Padova, Italy }
\affiliation {CAMK-PAN, 00-716 Warsaw, Poland }
\affiliation {Astronomical Observatory Warsaw University, 00-478 Warsaw, Poland }
\affiliation {University of Birmingham, Birmingham B15 2TT, United Kingdom }
\affiliation {Universit\`a degli Studi di Genova, I-16146 Genova, Italy }
\affiliation {INFN, Sezione di Genova, I-16146 Genova, Italy }
\affiliation {RRCAT, Indore MP 452013, India }
\affiliation {Faculty of Physics, Lomonosov Moscow State University, Moscow 119991, Russia }
\affiliation {SUPA, University of the West of Scotland, Paisley PA1 2BE, United Kingdom }
\affiliation {University of Western Australia, Crawley, Western Australia 6009, Australia }
\affiliation {Department of Astrophysics/IMAPP, Radboud University Nijmegen, P.O. Box 9010, 6500 GL Nijmegen, The Netherlands }
\affiliation {Artemis, Universit\'e C\^ote d'Azur, CNRS, Observatoire C\^ote d'Azur, CS 34229, Nice cedex 4, France }
\affiliation {MTA E\"otv\"os University, ``Lendulet'' Astrophysics Research Group, Budapest 1117, Hungary }
\affiliation {Institut de Physique de Rennes, CNRS, Universit\'e de Rennes 1, F-35042 Rennes, France }
\affiliation {Washington State University, Pullman, WA 99164, USA }
\affiliation {Universit\`a degli Studi di Urbino 'Carlo Bo', I-61029 Urbino, Italy }
\affiliation {INFN, Sezione di Firenze, I-50019 Sesto Fiorentino, Firenze, Italy }
\affiliation {University of Oregon, Eugene, OR 97403, USA }
\affiliation {Laboratoire Kastler Brossel, UPMC-Sorbonne Universit\'es, CNRS, ENS-PSL Research University, Coll\`ege de France, F-75005 Paris, France }
\affiliation {VU University Amsterdam, 1081 HV Amsterdam, The Netherlands }
\affiliation {University of Maryland, College Park, MD 20742, USA }
\affiliation {Center for Relativistic Astrophysics and School of Physics, Georgia Institute of Technology, Atlanta, GA 30332, USA }
\affiliation {Institut Lumi\`{e}re Mati\`{e}re, Universit\'{e} de Lyon, Universit\'{e} Claude Bernard Lyon 1, UMR CNRS 5306, 69622 Villeurbanne, France }
\affiliation {Laboratoire des Mat\'eriaux Avanc\'es (LMA), IN2P3/CNRS, Universit\'e de Lyon, F-69622 Villeurbanne, Lyon, France }
\affiliation {Universitat de les Illes Balears, IAC3---IEEC, E-07122 Palma de Mallorca, Spain }
\affiliation {Universit\`a di Napoli 'Federico II', Complesso Universitario di Monte S.Angelo, I-80126 Napoli, Italy }
\affiliation {NASA/Goddard Space Flight Center, Greenbelt, MD 20771, USA }
\affiliation {Canadian Institute for Theoretical Astrophysics, University of Toronto, Toronto, Ontario M5S 3H8, Canada }
\affiliation {Tsinghua University, Beijing 100084, China }
\affiliation {Texas Tech University, Lubbock, TX 79409, USA }
\affiliation {The Pennsylvania State University, University Park, PA 16802, USA }
\affiliation {National Tsing Hua University, Hsinchu City, Taiwan 30013, R.O.C. }
\affiliation {Charles Sturt University, Wagga Wagga, New South Wales 2678, Australia }
\affiliation {University of Chicago, Chicago, IL 60637, USA }
\affiliation {Caltech CaRT, Pasadena, CA 91125, USA }
\affiliation {Korea Institute of Science and Technology Information, Daejeon 305-806, Korea }
\affiliation {Carleton College, Northfield, MN 55057, USA }
\affiliation {Universit\`a di Roma 'La Sapienza', I-00185 Roma, Italy }
\affiliation {University of Brussels, Brussels 1050, Belgium }
\affiliation {Sonoma State University, Rohnert Park, CA 94928, USA }
\affiliation {Northwestern University, Evanston, IL 60208, USA }
\affiliation {University of Minnesota, Minneapolis, MN 55455, USA }
\affiliation {The University of Melbourne, Parkville, Victoria 3010, Australia }
\affiliation {The University of Texas Rio Grande Valley, Brownsville, TX 78520, USA }
\affiliation {The University of Sheffield, Sheffield S10 2TN, United Kingdom }
\affiliation {University of Sannio at Benevento, I-82100 Benevento, Italy and INFN, Sezione di Napoli, I-80100 Napoli, Italy }
\affiliation {Montclair State University, Montclair, NJ 07043, USA }
\affiliation {Universit\`a di Trento, Dipartimento di Fisica, I-38123 Povo, Trento, Italy }
\affiliation {INFN, Trento Institute for Fundamental Physics and Applications, I-38123 Povo, Trento, Italy }
\affiliation {Cardiff University, Cardiff CF24 3AA, United Kingdom }
\affiliation {National Astronomical Observatory of Japan, 2-21-1 Osawa, Mitaka, Tokyo 181-8588, Japan }
\affiliation {School of Mathematics, University of Edinburgh, Edinburgh EH9 3FD, United Kingdom }
\affiliation {Indian Institute of Technology, Gandhinagar Ahmedabad Gujarat 382424, India }
\affiliation {Institute for Plasma Research, Bhat, Gandhinagar 382428, India }
\affiliation {University of Szeged, D\'om t\'er 9, Szeged 6720, Hungary }
\affiliation {Embry-Riddle Aeronautical University, Prescott, AZ 86301, USA }
\affiliation {University of Michigan, Ann Arbor, MI 48109, USA }
\affiliation {Tata Institute of Fundamental Research, Mumbai 400005, India }
\affiliation {American University, Washington, D.C. 20016, USA }
\affiliation {University of Massachusetts-Amherst, Amherst, MA 01003, USA }
\affiliation {University of Adelaide, Adelaide, South Australia 5005, Australia }
\affiliation {West Virginia University, Morgantown, WV 26506, USA }
\affiliation {University of Bia{\l }ystok, 15-424 Bia{\l }ystok, Poland }
\affiliation {SUPA, University of Strathclyde, Glasgow G1 1XQ, United Kingdom }
\affiliation {IISER-TVM, CET Campus, Trivandrum Kerala 695016, India }
\affiliation {Institute of Applied Physics, Nizhny Novgorod, 603950, Russia }
\affiliation {Pusan National University, Busan 609-735, Korea }
\affiliation {Hanyang University, Seoul 133-791, Korea }
\affiliation {NCBJ, 05-400 \'Swierk-Otwock, Poland }
\affiliation {IM-PAN, 00-956 Warsaw, Poland }
\affiliation {Rochester Institute of Technology, Rochester, NY 14623, USA }
\affiliation {Monash University, Victoria 3800, Australia }
\affiliation {Seoul National University, Seoul 151-742, Korea }
\affiliation {University of Alabama in Huntsville, Huntsville, AL 35899, USA }
\affiliation {ESPCI, CNRS, F-75005 Paris, France }
\affiliation {Universit\`a di Camerino, Dipartimento di Fisica, I-62032 Camerino, Italy }
\affiliation {Southern University and A\&M College, Baton Rouge, LA 70813, USA }
\affiliation {College of William and Mary, Williamsburg, VA 23187, USA }
\affiliation {Instituto de F\'\i sica Te\'orica, University Estadual Paulista/ICTP South American Institute for Fundamental Research, S\~ao Paulo SP 01140-070, Brazil }
\affiliation {University of Cambridge, Cambridge CB2 1TN, United Kingdom }
\affiliation {IISER-Kolkata, Mohanpur, West Bengal 741252, India }
\affiliation {Rutherford Appleton Laboratory, HSIC, Chilton, Didcot, Oxon OX11 0QX, United Kingdom }
\affiliation {Whitman College, 345 Boyer Ave, Walla Walla, WA 99362 USA }
\affiliation {National Institute for Mathematical Sciences, Daejeon 305-390, Korea }
\affiliation {Hobart and William Smith Colleges, Geneva, NY 14456, USA }
\affiliation {Janusz Gil Institute of Astronomy, University of Zielona G\'ora, 65-265 Zielona G\'ora, Poland }
\affiliation {Andrews University, Berrien Springs, MI 49104, USA }
\affiliation {Universit\`a di Siena, I-53100 Siena, Italy }
\affiliation {Trinity University, San Antonio, TX 78212, USA }
\affiliation {University of Washington, Seattle, WA 98195, USA }
\affiliation {Kenyon College, Gambier, OH 43022, USA }
\affiliation {Abilene Christian University, Abilene, TX 79699, USA }

\begin{abstract}
The LIGO detection of the gravitational wave transient GW150914, from the inspiral and merger of two black holes with masses $\gtrsim\,30$\,M$_{\odot}$, suggests a population of
binary black holes with relatively high mass.
This observation implies that the stochastic gravitational-wave background from binary black holes, created from the incoherent superposition of all the merging binaries in the Universe, could be higher than previously expected.
Using the properties of GW150914, we estimate the energy density of such a background from binary black holes.
In the most sensitive part of the Advanced LIGO/Virgo band for stochastic backgrounds
(near $\unit[25]{Hz}$),
we predict $\Omega_\text{GW}(f=\unit[25]{Hz}) = \OmegaRange $ with 90\% confidence.
This prediction is robustly demonstrated for a variety of formation scenarios with different parameters.
The differences between models are small compared to the statistical uncertainty arising from the currently poorly constrained local coalescence rate.
We conclude that this background is potentially measurable by the Advanced LIGO/Virgo detectors operating at their projected final sensitivity.

\end{abstract}


\maketitle

\pagestyle{plain}

\noindent
{\em Introduction} ---
On September 14, 2015 the Advanced LIGO~\cite{cqg.27.084006.10,cqg.32.074001.15} Hanford and Livingston detectors observed the gravitational-wave event GW150914 with a significance in excess of \CBCEVENTSIGMA$\sigma$~\cite{gw150914}.
The observed signal is consistent with a binary black hole waveform with component masses of $m_1=$\MONESCOMPACT\, M$_\odot$ and $m_2=$\MTWOSCOMPACT\, M$_\odot$, as measured in the source frame, and coalescing at a luminosity distance of \DISTANCECOMPACT\, Mpc, corresponding to a redshift of $z = \REDSHIFTCOMPACT$ \cite{gw150914,gw150914PE}.

For every event like GW150914 observed by advanced gravitational-wave detectors, there are many more too distant to be resolved.  The gravitational waves from these unresolvable events combine to create a stochastic background, which can be detected by correlating the signals from two or more gravitational-wave detectors~\cite{1999PhRvD..59j2001A}. While it has long been known that the advanced detectors could observe such a background, the detection of GW150914 suggests that the binary black hole background level is likely to be at the higher end of previous predictions (see, e.g.,~\cite{2011RAA....11..369R,2011ApJ...739...86Z,2011PhRvD..84h4004R,2011PhRvD..84l4037M,2012PhRvD..85j4024W,2013PhRvD..87d2002W,2013MNRAS.431..882Z,2015A&A...574A..58K}). 

Heavy black holes like GW150914 are predicted to form in low-metallicity stellar environments, lower than about half of solar metallicity, and in the presence of relatively weak massive-star winds \cite{gw150914astro}. These masses are also larger than the masses inferred from reliable dynamical measurements in black-hole X-ray binaries. More massive binaries emit more energy in gravitational waves. Hence, the measurement of the component masses of GW150914 favors a higher amplitude of the corresponding gravitational-wave background.

In addition, the coalescence rate of binary black holes like GW150914 in the local Universe is estimated to be $\alphabetrateone$ \cite{gw150914rate} median with 90\% credible interva.
This rate excludes the lower end of pre-detection rate estimates \cite{gw150914astro}, while being consistent with the higher end.
A higher coalescence rate also implies a brighter stochastic background.

There are currently two possible formation channels that are consistent with the GW150914 event \cite{gw150914astro}. Binary black holes may be formed from isolated binaries of massive stars in galactic fields, or through dynamical interactions in dense stellar environments such as globular clusters  \cite{gw150914astro}.  The evolution of the merger rate with redshift depends in part on the assumed formation scenario.

In this paper we discuss the detectability of the
stochastic background produced by binary black holes throughout the Universe based on
the measured properties of GW150914.

\medskip
\noindent
{\em Binary black hole background} ---
The energy density spectrum of gravitational waves is described by the following dimensionless quantity \cite{1999PhRvD..59j2001A}:
\begin{eqnarray}
  \Omega_\text{GW}(f) = \frac{f}{\rho_c} \frac{d\rho_\text{GW}}{df}\,,
\end{eqnarray}
where  $d\rho_\text{GW}$ is the energy density in the frequency
interval $f$ to $f+df$, $\rho _{c} = 3H_0^2c^2/8\pi G$ is the critical energy density required to close the Universe, and $H_0=67.8 \pm 0.9{\rm \; km/s/Mpc}$ \cite{Ade:2015xua}.

A population of binary black holes is characterized by the distribution of the intrinsic source parameters $\theta$ (usually the component masses and spin). Since this distribution is unknown at present, following \cite{gw150914rate} and \cite{Kim2003} we divide the distribution into distinct classes corresponding to the observed candidates.
If binary black holes in some class $k$, with source parameters $\theta_k$, merge at a rate $R_m(z;\theta_k)$ per unit comoving volume $V_c$ per unit source time, then the total gravitational-wave energy density spectrum is given by (see, e.g.~\cite{2011RAA....11..369R,2011ApJ...739...86Z,2011PhRvD..84h4004R,2011PhRvD..84l4037M,2012PhRvD..85j4024W,2013PhRvD..87d2002W,2013MNRAS.431..882Z,2015A&A...574A..58K}):

\begin{equation}
  \Omega_{\rm{GW}}(f;\theta_k)=\frac{f}{\rho_c H_0} \int_0^{z_{\rm max}}  dz \frac{R_m(z,\theta_k) \frac{dE_{\rm{GW}}}{df_s}(f_s,\theta_k)}{(1+z) E(\Omega_{\rm M},\Omega_{\Lambda},z)} ,
\label{eq:omega_flux}
\end{equation}
and the total energy density spectrum is the sum of $\Omega_\text{GW}(f;\theta_k)$ from each class.\footnote{When the distribution of the source parameters is better understood after multiple detections, the discrete sum can be replaced by a continuous integral.}
In Eq.~\ref{eq:omega_flux}, $dE_{\rm GW}/df_s(f_s,\theta_k)$ is the spectral energy density of a source of class $k$ at the frequency $f_s=f(1+z)$, which depends on the source parameters $\theta_k$; $E(\Omega_{\rm M},\Omega_{\Lambda},z) = \sqrt{\Omega_{\rm M} (1+z)^3 + \Omega_{\Lambda}}$ captures the dependence of the comoving volume on redshift for the standard flat cosmology model, with $\Omega_{\rm M} = 0.31$ and $\Omega_{\Lambda} = 1-\Omega_{\rm M}$. The $(1+z)$ factor in the denominator of Eq.~\ref{eq:omega_flux} corrects for the cosmic expansion, converting time in the source frame to the detector frame. The parameter $z_{\rm max}$ corresponds to the time of the first coalescences. We set $z_{\max}=10$, noting, however, that sources above $z \sim 5$ contribute very little to the total background (see, e.g.,~\cite{2011RAA....11..369R,2011ApJ...739...86Z,2011PhRvD..84h4004R,2011PhRvD..84l4037M,2012PhRvD..85j4024W,2013PhRvD..87d2002W,2013MNRAS.431..882Z,2015A&A...574A..58K}).

The merger rate $R_m(z;\theta_k)$ is a convolution of the binary formation rate $R_f(z;\theta_k)$ with the distribution of the time delays $P(t_d;\theta_k)$ between binary black hole formation and merger (see e.g.,~\cite{2007PhR...442..166N}

\begin{equation}
R_m(z;\theta_k) = \int_{t_{\min}}^{t_{\max}} R_f(z_f;\theta_k)P(t_d;\theta_k) dt_d,
\end{equation}
where $t_d$ is the time delay, $z_f$ is the redshift at the formation time $t_f=t(z)-t_d$, and $t(z)$ is the age of the Universe at merger.

Inference on GW150914 \cite{gw150914PE}, along with expectations that gravitational-wave emission is efficient in circularizing the orbit \cite{gw150914astro}, allows us to restrict our models for $dE_{\rm{GW}}/df_s$ to circular orbits.  Measurements do not constrain the component spins in the orbital plane \cite{gw150914PE}; we therefore restrict our model to spins (anti-)aligned with the orbital angular momentum, and use the functional form of $dE_{\rm{GW}}/df_s$ derived in \cite{2011PhRvD..84h4037A}.
In addition to the component masses, this model depends on the effective spin parameter along the direction of the orbital angular momentum $\chi_{\mathrm{eff}}$, which takes values between $-1$ (in which both black holes have maximal spins anti-aligned with respect to the orbital angular momentum) and +1 (assuming maximally aligned spins) \cite{gw150914PE}.

\medskip
\noindent
{\em Fiducial Model} ---
The GW150914 event appears consistent with both the dynamic and field formation channels  \cite{gw150914astro}; however the field channel is currently better described in the stochastic background literature.
Thus our \Fiducial\ model is inspired by population synthesis studies of field binaries (see \cite{2015A&A...574A..58K}).

We assume that the binary black hole formation rate is proportional to the star formation rate (SFR) at metallicity $Z \leq Z_{\odot}/2$ \cite{gw150914astro}, where $Z_{\odot}$ is the solar metallicity. That is, to compute the binary black hole formation rate, the SFR is multiplied by the fraction of star formation occurring below the metallicity threshold $Z_c = Z_{\odot}/2$. 
For the SFR, we use the recent model \cite{2015MNRAS.447.2575V}, referred to here as ``Vangioni'', based on the gamma-ray burst rate of \cite{kistler} and on the normalization described in \cite{trenti,behroozi}. We adopt the mean metallicity--redshift relation of \cite{2014ARAA..52..415M}, rescaled upwards by a factor of 3 to account for local observations \cite{gw150914chris,2015MNRAS.447.2575V}. In addition, we assume the metallicity is $\log_{10}$-normally distributed with a standard deviation of 0.5 around the mean at each redshift \cite{2015MNRAS.452L..36D}.
We further assume that the time delay distribution follows
 $P(t_d) \propto t_d^\alpha$,
with $\alpha=-1$ for $t_d>t_{\min}$ \cite{2002ApJ...572..407B,2004JCAP...06..007A,2006ApJ...648.1110B,2006IJMPD..15..235D,2007ApJ...664.1000B,2007PhR...442..166N,2008ApJ...675..566O,2012ApJ...759...52D}, where $t_{\min}=\unit[50]{Myr}$ is the minimum delay time for a massive binary to evolve until coalescence \cite[e.g.,][]{dominik}, and a maximum time delay $t_{\max}$ equal to the Hubble time.  

The rest of the \Fiducial\ model parameters correspond to the median inferred parameters of GW150914: the chirp mass $M_c=\MCSOne$\,M$_{\odot}$, the symmetric mass ratio $\eta \sim0.25$, and the effective spin parameter $\chi_{\mathrm{eff}}=\spinOne$.
We normalize the overall merger rate so that the local merger rate at $z=0$ matches the most conservative median inferred rate, \alphabetrateone \cite{gw150914rate}.

\begin{figure*}[hbtp!]

  \includegraphics[trim={0cm 0cm 0 0},clip,width=3.5in]{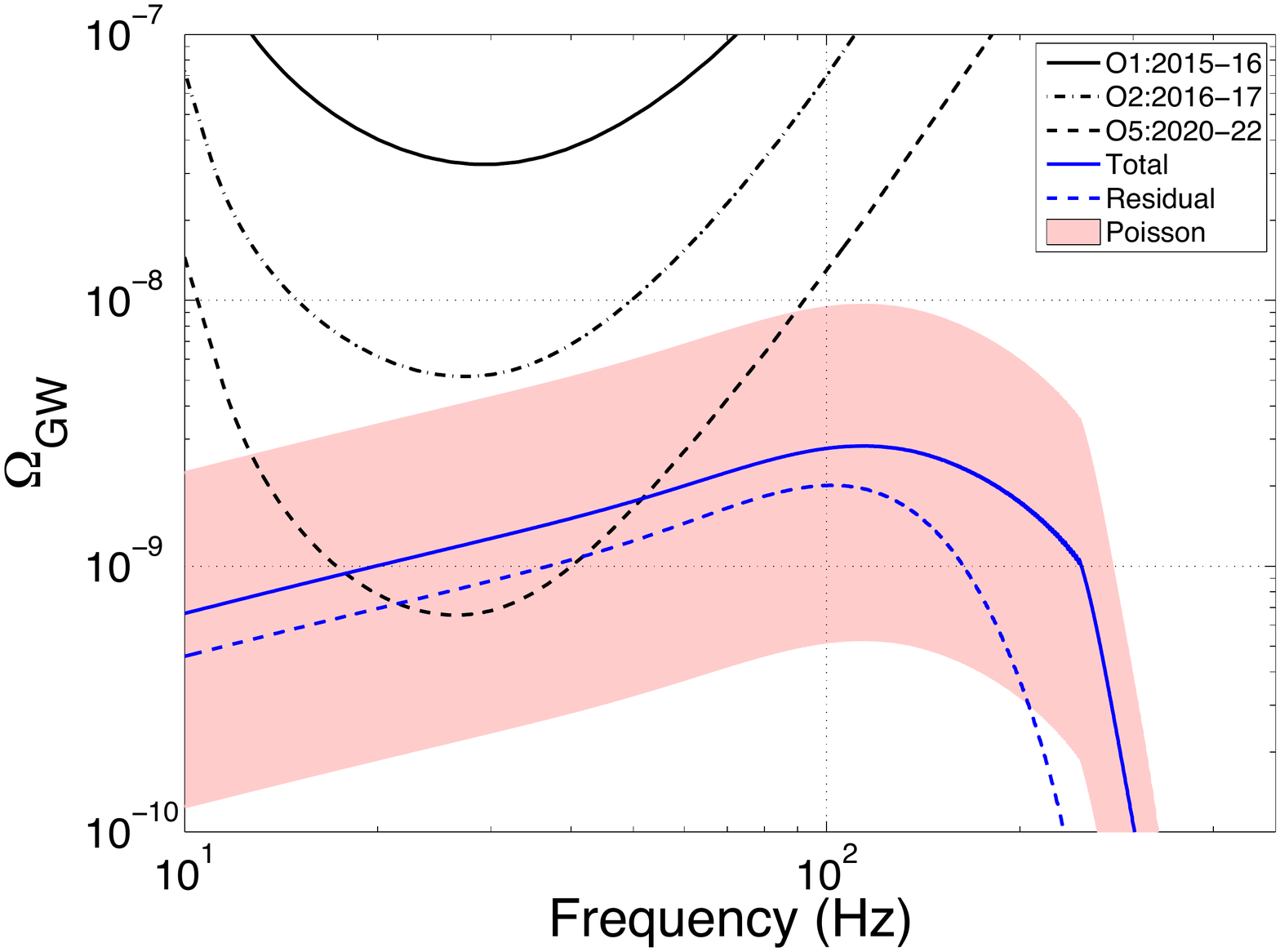}
  \includegraphics[trim={0cm 0cm 0 0},clip,width=3.5in]{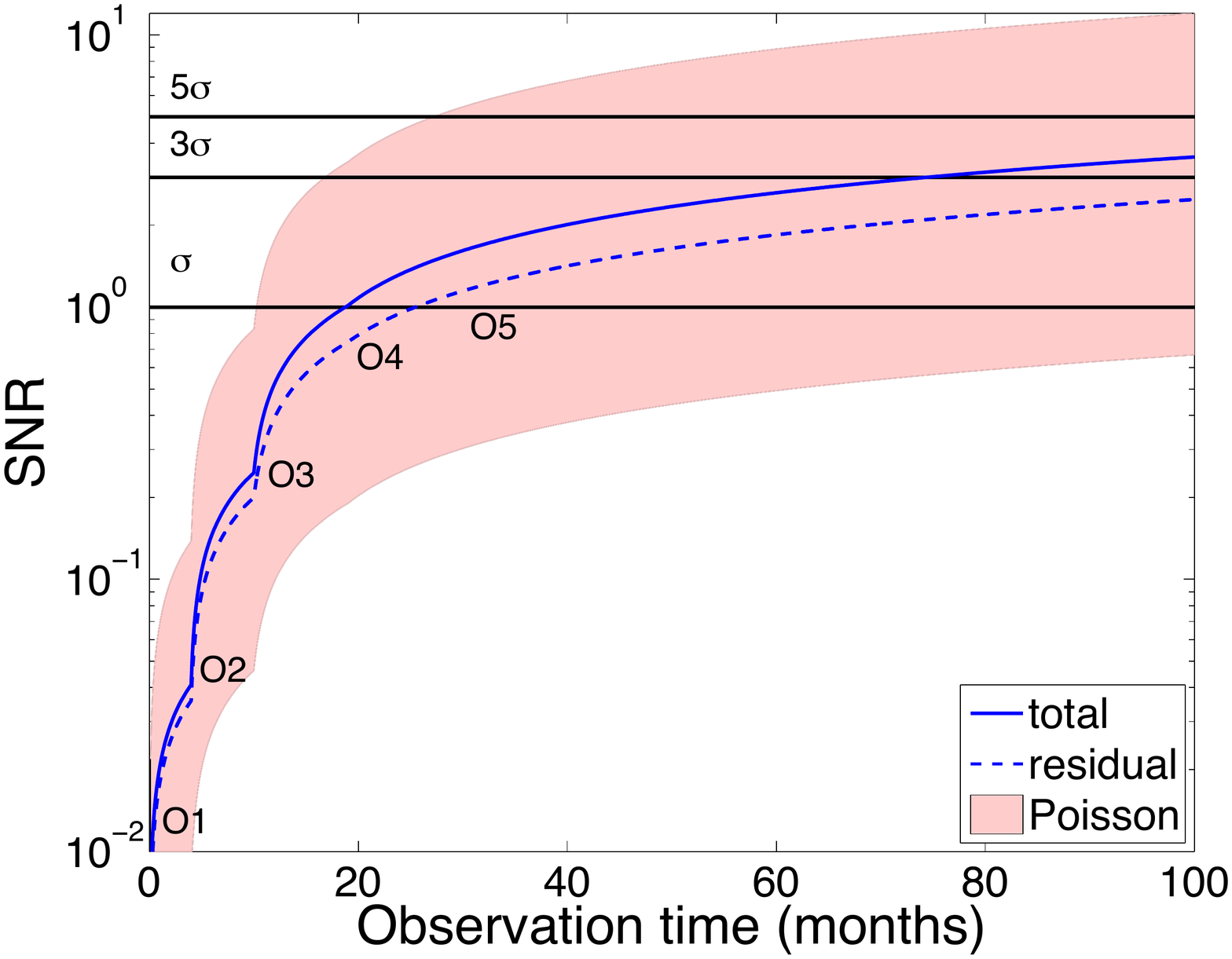}
  \caption{
    Expected sensitivity of the network of advanced LIGO and Virgo detectors to the \Fiducial\ field model.
     Left panel: Energy density spectra are shown in blue (solid for the total background; dashed for the residual background, excluding resolved sources, assuming final advanced LIGO and Virgo~\cite{cqg.27.084006.10,cqg.32.074001.15} sensitivity).
    The pink shaded region ``Poisson'' shows the 90\% CL statistical uncertainty, propagated from the local rate measurement, on the total background.
    The black power-law integrated curves show the $1\sigma$ sensitivity of the network expected for the two first observing runs O1 and O2, and for 2 years at the design sensitivity in O5.
    (O3 and O4 are not significantly different than O5; see Table~\ref{tab:phases}.)
    If the astrophysical background spectrum intersects a black line, it has expected $\text{SNR} \geq 1$.
    In both panels we assume a coincident duty cycle of 33\% for O1 (actual) and  50\% for all other runs (predicted).
    Right panel: Predicted SNR as a function of total observing time.
    The blue lines and pink shaded region have the same interpretation as in the left panel.
    Each observing run is indicated by an  improvement in the LIGO-Virgo network sensitivity~\cite{Aasi:2013wya}, which results in a discontinuity in the slope. 
    The thresholds for $\text{SNR}=1$, $3$ (false-alarm probability $<3\times 10^{-3}$) and $5$ (false-alarm probability $< 6 \times 10^{-7}$) are indicated by horizontal lines.
  \label{fig:fiducial}
  }
\end{figure*}

\medskip
\noindent
{\em Results} ---
We plot $\Omega_\text{GW}(f)$ for the \Fiducial\ model as a solid blue curve in Fig.~\ref{fig:fiducial}a.  The curve is shown against the pink shaded region, which represents the $90\%$ credible interval statistical uncertainty in the local rate. Considering this uncertainty, we predict $\Omega_\text{GW}(f=\unit[25]{Hz}) = \OmegaRange$.
The spectrum is well approximated by a power law $\Omega_{\rm{GW}} (f) \propto f^{2/3}$ at low frequencies where the contribution from the inspiral phase is dominant and the spectral energy density is $dE_{\rm GW}/df_s = [(G\pi)^{2/3}/3]M_c^{5/3}f_s^{-1/3}$. This power law remains a good approximation until the spectrum reaches a maximum at $f \sim 100$ Hz.  The shape is in agreement with previous predictions (see, e.g.,~\cite{2011ApJ...739...86Z,2011PhRvD..84h4004R,2011PhRvD..84l4037M,2012PhRvD..85j4024W,2013PhRvD..87d2002W,2013MNRAS.431..882Z,2015A&A...574A..58K}), except that the maximum is shifted to lower frequencies, due to the higher mass considered.

This calculation of $\Omega_\text{GW}(f)$ captures the total energy density in gravitational waves generated by binary black hole coalescences.
In practice, some of these sources will be individually detected as resolved binaries.
We define ``the residual background'' as the energy density spectrum that excludes potentially resolvable binaries.
While the total background is a property of the Universe, the residual background is detector-dependent.
As sensitivity improves, the surveyed volume increases, more binaries are resolved and the residual background decreases.

The dashed blue curve in Fig.~\ref{fig:fiducial}a represents the residual background calculated for the network of the Advanced LIGO~\cite{cqg.27.084006.10,cqg.32.074001.15} and Advanced Virgo~\cite{cqg.32.024001.15,jpcs.610.012014.15} detectors at final sensitivity, assuming that a binary black hole signal is detected if it is associated with a single-detector matched filter signal-to-noise ratio of $\rho>8$ in at least two detectors \cite{gw150914CBC}.
The difference between the two curves is about 30\% in the sensitive frequency band (10--50 Hz), indicating that the 
residual background carries complementary information about the binary black hole population.
Binaries with the same component masses as GW150914 can be detected at a redshift up to $z \lesssim 1.3$ by advanced detectors operating at design sensitivity if optimally located and oriented (see Fig.~4 of \cite{gw150914astro}).
However, most sources at $z\gtrsim 0.4$ will not be individually resolvable because of unfavorable location and orientation.

The sensitive frequency band of the Advanced LIGO-Virgo network to a gravitational-wave background produced by binary black holes is 10--50 Hz, where $\Omega_{\rm{GW}} \sim f^{2/3}$. It corresponds to more than 95\% of the accumulated sensitivity \cite{2015A&A...574A..58K,2013MNRAS.431..882Z,2015PhRvD..92f3002M}.
The black curves shown in Fig.~\ref{fig:fiducial}a are {\em power-law integrated curves}~\cite{locus}, 
which represent the expected $1\sigma$
sensitivity of the standard cross-correlation search~\cite{1999PhRvD..59j2001A}
to power-law gravitational-wave backgrounds, of which
the $\Omega_{\rm GW}(f) \propto f^{2/3}$ spectrum for binary inspirals is
an example.
A power-law integrated curve is calculated by taking the locus of
power-law spectra that have expected ${\rm SNR}=1$,
where~\cite{1999PhRvD..59j2001A}:

\begin{equation}
  \text{SNR} =\frac{3 H_0^2}{10 \pi^2} \sqrt{2T} \left[
\int_0^\infty df\>
\sum_{i=1}^n\sum_{j>i}
\frac{\gamma_{ij}^2(f)\Omega_{\rm GW}^2(f)}{f^6 P_i(f)P_j(f)} \right]^{1/2}\,,
\label{eq:snrCC}
\end{equation}
for a network of detectors $i=1,2,\cdots, n$.
Hence, if the spectrum of an astrophysical background intersects a black
curve, then it has an expected ${\rm SNR}\ge 1$.
In Eq. \ref{eq:snrCC}, $P_i(f)$ and $P_j(f)$ are the one-sided strain noise
power spectral densities of two detectors; $\gamma_{ij}(f)$ is the normalized
isotropic overlap reduction function
~\cite{1993PhRvD..48.2389F,1992PhRvD..46.5250C};
and $T$ is the accumulated coincident observation time.
While Eq.~\ref{eq:snrCC} is derived by assuming a Gaussian
background~\cite{1999PhRvD..59j2001A}, it can also be applied to non-Gaussian
backgrounds (with signals that are clearly separated in time) such as the
binary black hole background considered here~\cite{2014PhRvD..89h4063M}.
The different black curves shown in this plot illustrate the improvement
in expected sensitivity in the coming years.

\begin{table*}
\caption{
  Different phases in the evolution of the aLIGO--AdVirgo detector network over the next several years.
  The aLIGO and AdVirgo noise curves corresponding to high-sensitivity versions of ``Early'', ``Mid'', ``Late'', and  ``Design'' spectra are taken from \cite{Aasi:2013wya}.
  Note that AdVirgo did not participate in the O1 observing run, so is not included in the first phase. 
  ``Duration'' refers to the planned calendar time as opposed to the amount of accumulated data, for which we assume a duty cycle of $33\%$ for O1 (actual) and $50\%$ for all other runs (predicted).
\label{tab:phases}}
\begin{ruledtabular}
\begin{tabular}{l c c c c }

Observing run & Epoch & Duration (months) & aLIGO sensitivity & AdVirgo sensitivity\\
\hline
O1 & 2015--2016 & 4 & Early &---\\
O2 & 2016--2017 & 6 &Mid &Early\\
O3 & 2017--2018 & 9 & Late & Mid\\
O4 & 2019 & 12 & Design &Late\\
O5 & 2020+ & -- & Design &Design\\
\end{tabular}
\end{ruledtabular}
\label{tab:parameters}
\end{table*}

Following \cite{Aasi:2013wya,2015PhRvD..92f3002M}, we consider five different phases, denoted O1 to O5,
corresponding to the first five observing runs, summarized in Table ~\ref{tab:phases}.
For clarity, we show only the O1, O2, and O5 power-law integrated curves
since the differences between the projected sensitivities for O3, O4, and O5 are relatively small.
In Fig.~\ref{fig:fiducial}b, we plot the expected accumulated SNR for the \Fiducial\ model
as a function of total observation time.
For both the sensitivity curves and the accumulated SNR, we assume a coincident duty cycle for each pair of detectors of $33\%$ for O1 (actual) and $50\%$ for all other runs (predicted).
The total background associated with the \Fiducial\ model could be identified with ${\rm SNR}=3$, corresponding to false alarm probability $<3\times 10^{-3}$, after approximately 6 years of observing.
In the most optimistic scenario given by statistical uncertainties, the total background could
be identified after 1.5 years with ${\rm SNR}=3$ and after approximatively 2 years with ${\rm SNR}= 5$, which is even before design sensitivity is reached. It would take about 2 years of observing to achieve ${\rm SNR}=3$ and about 3.5 years for ${\rm SNR}=5$ for the optimistic residual background.
The most pessimistic case considered here is out of reach of the advanced detector network but is in the scope of third generation detectors.

\medskip
\noindent
{\em Alternative Models} ---
We now investigate the impact of possible variations on the \Fiducial\ model. We consider the following alternatives:
\begin{itemize}
\item \AltSFR~differs from the \Fiducial\ model in assuming a different SFR proposed by Tornatore et al.~\cite{2007MNRAS.382.1050T}, who combined observations and simulations at higher redshift; the formation rate is assumed to be proportional to the SFR, with no metallicity threshold. We also considered the Madau \& Dickinson SFR~\cite{2014ARAA..52..415M}, and found that it produces an energy density spectrum that is essentially indistinguishable from the \Fiducial\ model.
\item \LongDelay~is identical to the \Fiducial\ model but assumes a significantly longer minimum time delay $t_{\min}=\unit[5]{Gyr}$, potentially consistent with binary black hole formation via the chemically homogeneous evolution of rapidly rotating massive stars in very tight binaries \cite{MandelDeMink:2016}. 

\item \LowMetallicity~is the same as \Fiducial, but assumes that a significantly lower metallicity is required to form heavy black holes, with a threshold of  $Z_c=Z_{\odot}/10$ \cite{gw150914astro}.

\item \FlatDelay~assumes a flat time delay distribution, $\alpha=0$, with $t_{\min}=\unit[50]{Myr}$ and $t_{\max}=\unit[1]{Gyr}$.  This is inspired by the supposition that dynamical formation of the most massive binaries is likely to happen fairly early in the history of the host environment. 
\item \ConstRate~follows the assumption of \cite{gw150914} in considering a redshift-independent merger rate, $R_m(z) = \unit[\alphabetrateoneSIMPLE]{Gpc^{-3}yr^{-1}}$. 

\item \TwoEvents~is the same as the \Fiducial\ model except we add a second class of binary black hole sources corresponding to lower-mass systems with a smaller range for individual detections during O1. As an example, we assume a chirp mass of half the mass of GW150914, $M_c = 15$ M$_\odot$ (corresponding to the second most significant trigger in the compact binary coalescence search \cite{gw150914CBC}) and a local merger rate of 60 Gpc$^{-3}$ yr$^{-1}$, about 4 times larger than the rate of GW150914. We assume here that the metallicity threshold is $Z_c=$\,Z$_{\odot}$. 
\end{itemize}

\begin{figure}[hbtp!]
\includegraphics[width=3.5in]{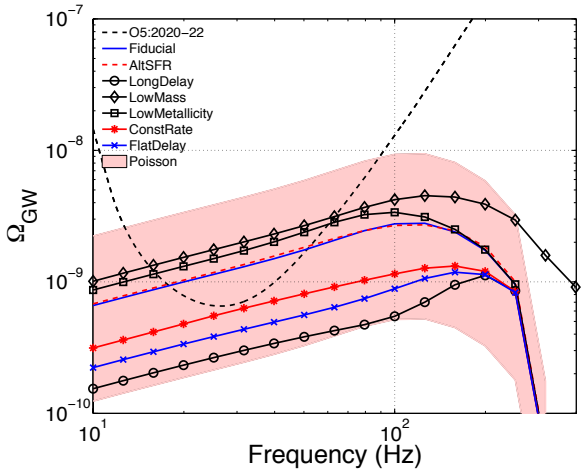}
  \caption{
    Energy density spectra for the different models summarized in the text. The pink shaded region ``Poisson'' shows the 90\% CL statistical uncertainty propagated from the local rate measurement, on the \Fiducial\ model. 
    The black dashed curve shows the design sensitivity of the network of Advanced LIGO ~\cite{cqg.27.084006.10,cqg.32.074001.15} and Virgo~\cite{cqg.32.024001.15,jpcs.610.012014.15}; see Tab.~\ref{tab:phases}.
    If the astrophysical background spectrum intersects with the dashed black line, it has expected $\text{SNR} \geq 1$. 
  \label{fig:models}
  }
\end{figure}

Figure~\ref{fig:models} shows the impact of alternative models described above.  The differences in the spectra of alternative models are not negligible.  However, all models considered here fall within the range of statistical uncertainty in the local merger rate estimate relative to the \Fiducial\ model in the sensitive frequency band.

The impact of an alternative star formation rate, as examined through model \AltSFR, is particularly small, indicating that the accuracy of SFR models is not a significant source of systematic error in predicting the strength of 
the gravitational-wave background.  

Relative to the \Fiducial\ model, the \LongDelay, \FlatDelay, and \ConstRate~models all predict fewer binaries at $z>0$, even though all of these models are constrained to have the same local merger rate ($z=0$). These latter three models consequently yield a lower energy density. 
The \LowMetallicity~model is characterized by a greater high-redshift merger rate than the \Fiducial\ model, with significant merger rates extending out to $z \sim 5-6$. This is because very little of the local Universe has the required low metallicity, so the local mergers come from the long time-delay tail of a large high-redshift population. Consequently, the \LowMetallicity~model has a higher overall normalization, as well as a different spectral shape at frequencies above 100 Hz due to the redshifting of the dominant high-$z$ contribution to the gravitational-wave background to lower frequencies.

Relative to the \Fiducial\ model, the \TwoEvents~model shows a greater energy density at all frequencies, particularly at high frequencies due to the signals from lower-mass binaries.  
This model indicates that if there is a
significant rate of mergers of binaries with smaller masses than GW150914, their contribution to the gravitational-wave energy density spectrum could be significant. The delta-function mass distributions assumed in all models are motivated by the observed candidates, but are not realistic. We have analyzed two alternative broad mass distributions considered in \cite{gw150914rate},
flat in the log-mass of the component black holes and a Salpeter-like mass function for the larger black hole with a flat mass ratio; these yield broadly consistent energy densities.
We have not carried out a systematic study of black hole spin.
Measurements of GW150914 prefer small values of spin in the direction of orbital momentum, but spins in the orbital plane are not constrained.
Preliminary studies suggest that $\Omega_\text{GW}(f)$ could change by a factor of $\lesssim 2$ for models including spin.

\medskip
\noindent
{\em Conclusions and discussion} ---
The detection of gravitational waves from GW150914 is consistent with the existence of
high-mass binary black hole mergers with a coalescence rate of tens per Gpc$^3$ per year.
As a consequence, the stochastic background from binary black holes is expected to be at the higher end of previous predictions (see, e.g.,~\cite{2011ApJ...739...86Z,2011PhRvD..84h4004R,2011PhRvD..84l4037M,2012PhRvD..85j4024W,2013PhRvD..87d2002W,2013MNRAS.431..882Z,2015A&A...574A..58K}).
We have shown that, for the \Fiducial\ field model, the energy density spectrum is $\Omega_\text{GW}(f=\unit[25]{Hz}) = 1.1_{-0.9}^{+2.7} \times 10^{-9}$ with 90\% confidence.
This, in turn, implies that the background may be measured by the network of advanced LIGO and Virgo detectors operating at or near their final sensitivity.
The uncertainty in this prediction arises from the statistical uncertainty in the local merger rate estimate.

Our predictions are subject to statistical fluctuations in the observed $\Omega_\text{GW}(f)$ due to random realizations of the binaries that coalesce during the observing run.
These fluctuations are much smaller than the current local merger uncertainty~\cite{2014PhRvD..89h4063M}.
The predictions may also be conservative. Throughout, we have assumed the use of the standard cross-correlation statistic, which is known to be sub-optimal for non-Gaussian backgrounds~\cite{DrascoFlanagan}.
The development of more sensitive non-Gaussian pipelines may hasten the detection of the binary black hole background~\cite{CornishRomano,lionel,popcorn}.

We have examined several alternative models for the merger rate evolution with redshift, representative of the uncertainties in the formation channels for high-mass binary black holes.
We find that all of these variations lie within the envelope of the uncertain local rate normalization in the 10--50 Hz band, as illustrated in Fig.~\ref{fig:models}. The power-law slope of the spectrum in this frequency band is not expected to deviate from $2/3$ unless there is a significant contribution from sources with high total mass merging at high redshift, $M(1+z) \gtrsim 200 M_\odot$. This illustrates the robustness of the predicted amplitude and power-law slope of the energy density spectrum.

However, this also implies that the stochastic background measurement with Advanced LIGO and Virgo detectors can only constrain the amplitude of the background power law in the 10--50 Hz sensitive frequency band.  The sensitivity of this search at the $2\sigma$ level will correspond to $\Omega_\text{GW} \sim 10^{-9}$ at 25 Hz with the full-sensitivity network of the Advanced LIGO/Virgo detectors.  Therefore, the stochastic search alone will not be able to distinguish between different model variations that have a similar effect on the spectrum in the 10-50 Hz band.  Future measurements of individual binary coalescences will help break at least some of these degeneracies, by providing a better estimate of the local merger rate and chirp mass distribution. Combining the two types of measurements (stochastic and individual coalescence event) could therefore help distinguish between different astrophysical formation scenarios for binary black holes~\cite{stoch_paramest}, but the full potential of this approach may only be reached using third generation of gravitational-wave detectors.

\medskip
\noindent

{\em Acknowledgments}  ---
The authors gratefully acknowledge the support of the United States
National Science Foundation (NSF) for the construction and operation of the
LIGO Laboratory and Advanced LIGO as well as the Science and Technology Facilities Council (STFC) of the
United Kingdom, the Max-Planck-Society (MPS), and the State of
Niedersachsen/Germany for support of the construction of Advanced LIGO 
and construction and operation of the GEO600 detector. 
Additional support for Advanced LIGO was provided by the Australian Research Council.
The authors gratefully acknowledge the Italian Istituto Nazionale di Fisica Nucleare (INFN),  
the French Centre National de la Recherche Scientifique (CNRS) and
the Foundation for Fundamental Research on Matter supported by the Netherlands Organisation for Scientific Research, 
for the construction and operation of the Virgo detector
and the creation and support  of the EGO consortium. 
The authors also gratefully acknowledge research support from these agencies as well as by 
the Council of Scientific and Industrial Research of India, 
Department of Science and Technology, India,
Science \& Engineering Research Board (SERB), India,
Ministry of Human Resource Development, India,
the Spanish Ministerio de Econom\'ia y Competitividad,
the Conselleria d'Economia i Competitivitat and Conselleria d'Educaci\'o, Cultura i Universitats of the Govern de les Illes Balears,
the National Science Centre of Poland,
the European Commission,
the Royal Society, 
the Scottish Funding Council, 
the Scottish Universities Physics Alliance, 
the Hungarian Scientific Research Fund (OTKA),
the Lyon Institute of Origins (LIO),
the National Research Foundation of Korea,
Industry Canada and the Province of Ontario through the Ministry of Economic Development and Innovation, 
the Natural Science and Engineering Research Council Canada,
Canadian Institute for Advanced Research,
the Brazilian Ministry of Science, Technology, and Innovation,
Russian Foundation for Basic Research,
the Leverhulme Trust, 
the Research Corporation, 
Ministry of Science and Technology (MOST), Taiwan
and
the Kavli Foundation.
The authors gratefully acknowledge the support of the NSF, STFC, MPS, INFN, CNRS and the
State of Niedersachsen/Germany for provision of computational resources.

\end{document}